\documentclass{article} 
\usepackage{InspireMusic,times}
\usepackage{hyperref}
\usepackage{url}
\usepackage{amsmath,graphicx}
\usepackage{booktabs}
\usepackage{multicol, multirow, vwcol}
\usepackage{wrapfig, booktabs, lipsum}
\usepackage{tabularx, makecell, enumitem}
\usepackage{amssymb}
\usepackage{adjustbox}
\usepackage{pifont}

\usepackage{algorithm}
\usepackage{algorithmic}
\usepackage{amsmath}
\usepackage{amssymb}
\title{InspireMusic: Integrating Super Resolution and Large Language Model for High-Fidelity Long-Form Music Generation}
\author{Tongyi Lab, Alibaba Group}

\iclrfinalcopy 

\begin{document}\maketitle
\begin{abstract}
We introduce \textbf{InspireMusic}\footnote{The full list of authors and acknowledgments is presented at the end of this document. Correspondence regarding this technical report should be directed to \{chong.zhang, yukun.ma\}@alibaba-inc.com.}, a framework \textbf{in}tegrated \textbf{s}u\textbf{p}er resolut\textbf{i}on and la\textbf{r}ge languag\textbf{e} model for high-fidelity long-form \textbf{music} generation. A unified framework generates high-fidelity music, songs, and audio, which incorporates an autoregressive transformer with a super-resolution flow-matching model. This framework enables the controllable generation of high-fidelity long-form music at a higher sampling rate from both text and audio prompts. Our model differs from previous approaches, as we utilize an audio tokenizer with one codebook that contains richer semantic information, thereby reducing training costs and enhancing efficiency. This combination enables us to achieve high-quality audio generation with long-form coherence of up to $8$ minutes. Then, an autoregressive transformer model based on Qwen 2.5 predicts audio tokens. Next, we employ a super-resolution flow-matching model to generate high-sampling rate audio with fine-grained details learned from an acoustic codec model. Comprehensive experiments show that the InspireMusic-1.5B-Long model has a comparable performance to recent top-tier open-source systems, including MusicGen and Stable Audio 2.0, on subjective and objective evaluations. Readers explore more on the online demo\footnote{https://modelscope.cn/studios/iic/InspireMusic}\footnote{https://huggingface.co/spaces/FunAudioLLM/InspireMusic}. The code and pre-trained models are released\footnote{https://github.com/FunAudioLLM/InspireMusic}.
\end{abstract}
\vspace{-0.6cm}\section{Introduction} 
Music generation emerges as one of the most dynamic and rapidly evolving fields within artificial intelligence (AI), significantly contributing to the development of artificial intelligence-generated content. Traditionally, music composition has been a deeply creative process relying on human composers' skill, intuition, and emotional expression. However, recent advances in AI have transformed this process, enabling controllable music generation that mimics aspects of human creativity and offers new possibilities for musical composition. 

Earlier systems used primarily symbolic representations, such as MIDI~(\cite{engel2017neural}), which facilitate fine-grained control over the musical structure to generate music with rich timbre, precise music notes, high audio quality, and expressive nuances of real-world audio. In contrast, transformer-based models (e.g., MuLan~(\cite{Huang2022MuLanAJ}), BERT-based models (e.g., MusicLM~(\cite{agostinelli2023musiclm}), vector quantization based variational autoencoder (VQ-VAE) based models (e.g., Jukebox~\cite{dhariwal2020jukebox}) have improved audio generation to generate music with different styles, in a user-friendly way. However, generating long-form musical compositions that maintain coherence and fidelity over extended durations remains a fundamental challenge in this field. \\ Notable transformer-based models such as Jukebox~(\cite{dhariwal2020jukebox}), and MusicGen~(\cite{musicgen}) make significant strides by introducing autoregressive transformers to model musical structure. Jukebox, for instance, utilizes a multi-scale audio compression and decompression pipeline. MusicGen, which operates on discrete audio tokens using Encodec~(\cite{defossez2023encodec}) with multiple codebooks, simplifies the autoregressive process but is limited by short output lengths around $30s$ and at sampling rates of $32000Hz$. These models set benchmarks for music generation, but still have problems generating audio with long-range dependencies. \\ Diffusion- and flow-based methods, such as Stable Audio 2.0~(\cite{stableaudio2, stableaudioopen}), AudioLDM~(\cite{liu2023audioldm}), AudioLDM2~(\cite{Liu2023AudioLDM2L}), MusicFlow~(\cite{KR2024MusicFlowCF}), Jen-1~(\cite{Li2023Jen1}), emerge as promising alternatives, achieving impressive audio fidelity through iterative diffusion denoising processes. However, these methods often sacrifice computational efficiency and in some cases hard to maintain long-term structural coherence in the generated music. Specifically, Stable Audio 2.0 receives criticism for producing monotonous, repetitive melodies that lack musical diversity, and diffusion models generally suffer from a lack of fine-grained control when working with long and complex text prompts and intricate musical structures. According to~(\cite{wang2024diverse}), autoregressive models are good at memorization learned from discrete token sequences to generate sequences, which helps to generate music that closely resembles existing compositions and to fill in missing sections in a rule-consistent manner for the tasks such as text-to-music generation, music continuation, and music inpainting. In contrast, diffusion models tend to produce outputs with stronger structural characteristics.

Despite these advances, a substantial gap remains in music generation research. Namely, the ability to generate high-fidelity long-form music with precise control over global musical structure and local acoustic details. Although existing models excel in one aspect, e.g., musical structure or audio fidelity, no single framework has been able to effectively combine these elements with proper control over the generation process. In response to these challenges, we introduce InspireMusic, a unified framework designed to bridge the gap between high-fidelity raw audio generation and long-form music generation. Our approach integrates several cutting-edge techniques to overcome the limitations of prior models. First, we employ an ultra-low bitrate audio tokenization scheme: a $75Hz$ audio tokenizer~(\cite{ji2024wavtokenizer}) with one codebook that captures the global musical structure from audio and facilitates fast training and inference of the autoregressive model. We then employ a flow-matching-based super-resolution model for enhanced temporal coherence as well as higher fidelity and acoustic details. Specifically, we map the audio tokens generated from large language models to high-resolution fine-grained acoustic features obtained from audio with a higher sampling rate. This framework allows the generation of long-form, high-fidelity music with proper control over global structure and local acoustic details.

The contributions of this work are as follows. \begin{itemize} \item We introduce a unified framework that incorporates audio tokenization, autoregressive transformer, and super-resolution flow matching model to generate controllable high-fidelity audio with long-form coherence of up to $8$ minutes currently. \item We utilize a high-bitrate compression audio tokenizer with one codebook that preserves rich semantic information, thereby enabling efficient training and inference with large language models while generating long-form coherence. \item We also make this framework flexible, operating effectively in configurations that use LLMs alone or incorporate them with a flow-matching-based super-resolution model for generating audio of improved fidelity. \end{itemize}
\vspace{-0.8cm}
\section{Related Work} 
\label{sec:literature_review} 
Generative models for music and audio generation develop in multiple directions. In this section, we review several recent works that inspire the development of our approach.
\vspace{-0.5cm}
\subsection{Autoregressive Transformer Models}
Autoregressive (AR) transformer models play an important role in long-form sequence generation. MusicLM~(\cite{agostinelli2023musiclm}) pioneers the use of hierarchical AR transformers for raw audio generation by compressing audio into discrete tokens and then generating sequences with a transformer. This approach captures long-term musical structure but requires a complex, multi-stage decoding process. MusicGen~(\cite{musicgen}) builds upon this foundation by streamlining token-based generation, directly conditioning on text to produce music. While these models demonstrate the ability to generate convincing short segments, they often struggle with maintaining coherence over extended durations, leading to repetitive or meandering outputs.
\vspace{-0.5cm}
\subsection{Diffusion Models} 
Diffusion-based methods emerge as a promising alternative for high-fidelity audio generation. Stable Audio 2.0 (\cite{stableaudio2,stablemusic,stableaudioopen}) employs latent diffusion to iteratively denoise audio representations, achieving impressive perceptual quality. Similarly, AudioLDM~(\cite{liu2023audioldm}) leverages latent diffusion conditioned on text, offering an efficient pathway to generating complex audio. Despite their strengths in quality, diffusion models typically require many iterative steps during inference, which hampers real-time applications and sometimes disrupts the global structure of generated music.

\vspace{-0.5cm}
\subsection{Flow-matching Models} 
A recent trend is the use of flow-matching (FM) as an alternative to diffusion. Unlike traditional diffusion which simulates a gradual denoising process, FM directly learns a continuous mapping between noise and data distributions. This results in faster sampling and reduced complexity during inference. Our work builds upon these advances by incorporating a super-resolution flow-matching model to upscale audio tokens while preserving musical semantics. CosyVoice~(\cite{du2024cosyvoice}) and CosyVoice2~(\cite{du2024cosyvoice2}) use flow matching to map discrete tokens into speech waveforms. By training a model to follow the optimal “flow” from a simple noise distribution to the complex distribution of audio, FM achieves rapid convergence and high-quality generation. While still nascent, FM demonstrates its potential in domains like image generation and is now explored for audio. Our work leverages a super-resolution variant of flow-matching to upscale coarse token sequences to high-resolution audio, combining the benefits of efficient sampling with detailed output synthesis.
\vspace{-0.5cm}
\subsection{Music Generation} Recent works, such as Seed-Music~(\cite{seedmusic2024}) incorporate AR transformer and diffusion models to build controllable music generation systems with text encoder, audio tokenizer, and MIDI encoder, trained with the concatenated tokens from those encoders with labeled data, controlled by temporal conditioning on diffusion transformer. Seed-Music presents a unified framework that adapts to the evolving workflows of musicians, leveraging both AR language modeling and diffusion approaches to support controlled music generation and post-production editing. This adaptability allows for fine-grained style control and high-quality music production. While, Stable Audio 2.0~(\cite{stablemusic}) is a diffusion-based model combining diffusion U-Net with VAE decoder to generate high-fidelity long-form music with enhanced coherence and diversity, controlled by text embeddings obtained from CLAP~\footnote{https://github.com/LAION-AI/CLAP} pre-trained model and time embeddings.

The diverse approaches outlined above highlight a key insight: achieving both high-fidelity and long-term structural coherence in music generation requires integrating multiple generative paradigms. Autoregressive models excel at capturing long-form musical structures but face challenges with high fidelity. Diffusion and flow matching techniques offer high-quality audio but often require careful tuning to maintain global coherence. By combining these methods, InspireMusic overcomes these limitations and introduces a novel approach for generating long-form, high-fidelity, controllable music. This synthesis of techniques drives our hybrid architecture, enabling the creation of high-fidelity compositions while maintaining the flexibility to accommodate a variety of musical genres.
\vspace{-0.5cm}
\section{InspireMusic} \label{sec: framework}
\begin{figure*}[t]
    \centering
    \includegraphics[width=\linewidth]{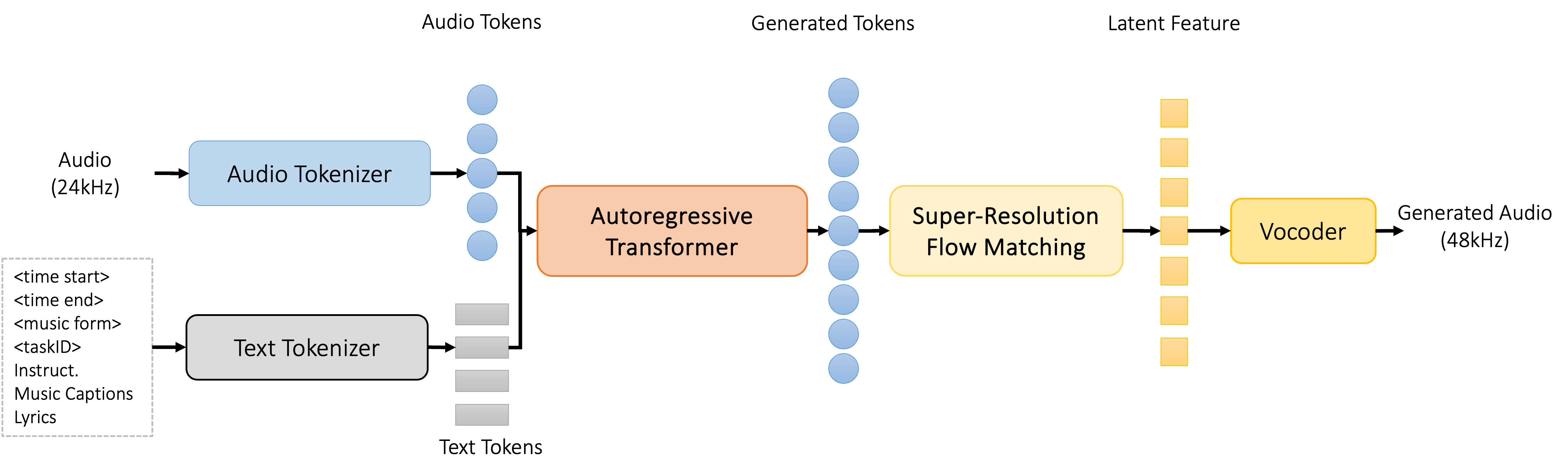}
    \caption{The overview of \textbf{InspireMusic} framework. InspireMusic is composed of a \textbf{audio tokenizers}, an \textbf{autoregressive transformer}, and a \textbf{super-resolution flow-matching} model. (1) Audio waveform of lower sampling rate has converted to discrete tokens via a high bitrate compression audio tokenizer. (2) The audio and text tokens are the inputs of an autoregressive model with the next token prediction to generate tokens. (3) Then the flow-matching model maps the generated tokens to the latent features with high-resolution fine-grained acoustic details obtained via Hifi-Codec~(\cite{yang2023hificodec}) from a higher sampling rate of audio to ensure the acoustic information flow connected with high fidelity through models. (4) The vocoder decoder then produces high-quality $48kHz$ audio.}
    \label{fig:inspiremusic}
\end{figure*}
The InspireMusic framework consists of three principal components: audio tokenizers, an autoregressive transformer, and a super-resolution flow-matching model. Figure~\ref{fig:inspiremusic} presents the overview of the InspireMusic framework.

\subsection{Audio Tokenization}
The first step in the InspireMusic framework is to convert the raw audio waveform into discrete audio tokens that can be efficiently processed and trained by the autoregressive transformer model. We use WavTokenizer as an audio tokenizer in InspireMusic.

WavTokenizer~(\cite{ji2024wavtokenizer}) serves as an audio tokenizer that compresses $24kHz$ audio into discrete tokens at a $75Hz$ token rate. WavTokenizer functions as an efficient audio tokenizer with only one codebook at $0.9kbps$ bandwidth while containing rich semantic information. It converts $24kHz$ audio into discrete tokens at a token rate of $75Hz$. These tokens capture the coarse audio information, and global musical structure, reducing the sequence length for efficient autoregressive transformer training. This is facilitated by designing a broader Vector Quantization (VQ) space, extending contextual windows, improving attention networks, and introducing a multi-scale discriminator along with an inverse Fourier transform (iFFT) in the decoder.

\subsection{Autoregressive Transformer}
A core of InspireMusic is an autoregressive (AR) transformer~(\cite{Vaswani2017AttentionIA}), with the backbone large language model of Qwen 2.5~(\cite{qwen2.5}) model series. The model learns to predict the next audio token in the sequence given the preceding tokens, to generate a long sequence of coarse audio tokens and global musical structure with long-form coherence. The audio tokens are extracted by an efficient audio tokenizer, i.e., WavTokenizer, which improves the model training process to learn to generate music with both temporal coherence and diversity.

The transformer trains using a next-token prediction objective, where the model generates a sequence conditioned on the input text descriptions, captions, tags ($s_{t}$), timestamps including time start ($ts$) and time end ($te$), music structures ($s$), label($l$), and audio tokens ($s_{a}$), as $S = \{s_{t}^1, s_{t}^2, \cdots, s_{t}^m, s_{ts}, s_{te}, s_{s}, s_{l}, s_{a}^1, s_{a}^2, \cdots, s_{a}^n\}$, where $T=m+n+4$. Train the AR transformer to effectively learn the long-term dependencies and ensures that the generated sequence adheres to the intended musical genre, descriptions, timestamps, and structures. Our experiments indicate that this module is capable of generating coherent musical compositions over extended durations. The input dimension sizes of 0.5B and 1.5B models are $896$ and $1536$, respectively.

The choice of Qwen 2.5 as the backbone language model is motivated by the performance, flexibility, and scalability of the model. Its design suits sequential generation tasks, such as music, where long-range dependencies must be captured between tokens. The use of a large language model in audio generation allows InspireMusic to leverage techniques from natural language processing, enabling it to generate coherent, structured, and meaningful music compositions from text or audio prompts.

In this work, we develop a series of models based on Qwen 2.5 with different parameter sizes. The InspireMusic-0.5B model is based on the Qwen2.5-0.5B model, whereas the more advanced variants, InspireMusic-1.5B and InspireMusic-1.5B-Long, rely on the Qwen2.5-1.5B model. InspireMusic has capacity to learn long-form coherence of the music and structural patterns in music for music generation tasks.

Classifier-free guidance (CFG)~(\cite{ho2022cfg}) proves effective in improving the generation quality of generative models. Therefore, we adapt the CFG into the AR transformer model. During training, we randomly drop the conditions with a fixed probability of $0.7$, enabling the AR model to learn both conditional and unconditional distributions. During inference, CFG also applies to the outputs of the AR model, and we recommend using a guidance scale of $3.0$. During the decoding process, the top-K sampling method samples the generated tokens with the default value of $350$.

\subsection{Super-Resolution Flow-Matching}
In the audio domain, super-resolution refers to the process of upscaling audio with a low sampling rate to a higher sample rate while preserving information in low-frequency components and enhancing fine-grained details in high-frequency components. In the context of audio generation, this typically means taking audio inputs sampled at a lower sampling rate (e.g., $24kHz$) and transforming them into higher-quality audio at a higher resolution (e.g., $48kHz$). Super-resolution in this setting aims to enhance the perceptual quality of audio while maintaining or improving the structural integrity of the original sound.

We propose a Super-Resolution Flow-Matching (SRFM) model to enhance low-resolution coarse audio tokens to high-resolution fine-grained audio outputs by learning optimal transformation paths between distributions. Unlike traditional iterative methods, SRFM employs flow matching techniques to directly model the mapping from coarse audio tokens from low sampling rate audio waveforms to fine-grained high-resolution latent audio features extracted from audio with a higher sampling rate (i.e., $48kHz$) via a $150Hz$ Hifi-Codec model, effectively capturing the underlying data distribution. This approach demonstrates the ability to generate high-fidelity, high-resolution outputs from low-resolution inputs in many domains.

For the $150Hz$ Hifi-Codec model, given a single channel audio sequence $X$ with the duration of $D$ as the inputs, an Encoder network $E$ takes the raw audio inputs and transforms them into hidden features $H$, a group residual quantization layer $Q$ with the codebook size of $4$ and each codebook dimension of $C$, and a decoder $G$ that reconstruct the audio signal from the compressed latent features, where in this study $H=1024$ and $C=1024$. 

SRFM models generate high-resolution outputs in a single pass by sampling from data for multiple iterative steps. By learning the optimal transformation paths between distributions, SRFM models produce outputs that closely resemble the true data distribution, resulting in high-quality, realistic images and audio. After generating a coarse sequence of tokens, the SRFM transforms these tokens into high-fidelity latent representations. Unlike standard diffusion models that require many iterative refinement steps, our SRFM model directly learns a continuous mapping from the discrete token space to the high-resolution latent space from $48kHz$ audio. The SRFM model minimizes a loss function based on the discrepancy between the predicted latent features and the latent features from high sampling rate audio. The SRFM effectively bridges the gap between the token sequence generated by the autoregressive transformer from lower sampling rate audio and the high-fidelity output expected in modern music production. By matching the flow of information across different sampling rates, this model enables InspireMusic to generate longer, more complex musical compositions without sacrificing audio quality. The final stage involves a vocoder that decodes the high-resolution latent features into a raw waveform. By integrating the Hifi-Codec~(\cite{yang2023hificodec}) with the SRFM output, the vocoder synthesizes $48kHz$ audio that remains both perceptually rich and free of artifacts. A dedicated vocoder decodes the high-resolution latent features into a raw waveform, producing perceptually rich, artifact-free audio.
\vspace{-0.5cm}
\subsection{Model Variants}
\begin{table}[t]
\centering
\caption{Overview of InspireMusic model configurations. This table presents the comparative configuration of various InspireMusic setups that integrate distinct combinations of autoregressive transformers and flow-matching models, with and without super-resolution, to support diverse music generation tasks from text-to-music synthesis to music continuation. The table also delineates configurations that without labeling the output sampling rate in the model name, the default output sampling rate is $48kHz$, thereby highlighting the framework’s versatility in handling multi-modal inputs and accommodating varying audio fidelity requirements.}
\footnotesize
\label{tab: models}
\begin{tabular}{l|c|c|c|c|c|l}
\toprule
Model Name                      & LLM & \begin{tabular}[c]{@{}c@{}}Flow\\Matching\end{tabular} & \begin{tabular}[c]{@{}c@{}}Super\\Resolution\end{tabular} &\begin{tabular}[c]{@{}c@{}}Long\\Form\end{tabular} & \begin{tabular}[c]{@{}c@{}}Output\\Sample\\Rate(Hz)\end{tabular} & Parameters \\ \midrule
InspireMusic-0.5B (w/o flow)      & \checkmark   &    &                  &           & $24000$                                                                     & $0.5B$       \\ 
InspireMusic-0.5B-24kHz           & \checkmark   & \checkmark &                  &           & $24000$                                                                     & $0.8B$  \\ 
InspireMusic-0.5B                 & \checkmark   & \checkmark  & \checkmark               &           & $48000$                                                                     & $0.8B$  \\ 
InspireMusic-1.5B (w/o flow)      & \checkmark   &    &                  &           & $24000$                                                                    & $1.5B$       \\ 
InspireMusic-1.5B-24kHz           & \checkmark   & \checkmark  &                  &           & $24000$                                                                     & $1.8B$  \\ 
InspireMusic-1.5B                 & \checkmark  & \checkmark  & \checkmark             &           & $48000$                                                                    & $1.8B$  \\ 
InspireMusic-1.5B-Long (w/o flow) & \checkmark   &    &                  & \checkmark         & $24000$                                                                     & $1.5B$       \\ 
InspireMusic-1.5B-Long            & \checkmark   & \checkmark  & \checkmark                & \checkmark         & $48000$                                                                    & $1.8B$  \\ \bottomrule
\end{tabular}
\end{table}
We train multiple variants of InspireMusic as listed below.
\textbf{InspireMusic-0.5B}: A lightweight model for 30-second compositions, based on the Qwen2.5-0.5B model~\footnote{https://huggingface.co/Qwen/Qwen2.5-0.5B} with a balanced performance between speed and quality.

\textbf{InspireMusic-1.5B}: A larger model for improved style control and quality on short pieces. Built on the Qwen2.5-1.5B model, this version generates 30-second music pieces and provides enhanced control over the style and structure of the output.

\textbf{InspireMusic-1.5B-Long}: Optimized for generating long-form compositions (up to 8 minutes) with enhanced structural coherence. Also based on the Qwen2.5-1.5B model, it ensures the model with long-form music while maintaining structural coherence and diversity. The previous model was trained with 30-second audio segment data, this model is capable of preserving the long-form coherence of music.

\subsection{Model Training and Inference}
The training procedure of InspireMusic includes training audio tokenizers, the autoregressive transformer model, and the flow-matching model.

\textbf{Training of Audio Tokenizers}: The audio tokenizer and music tokenizer train from scratch with music datasets sampled at $24kHz$ and $48kHz$, respectively. This approach enables the models to effectively process and generate high-fidelity audio across different sampling rates.

\textbf{Autoregressive Transformer Training}: The autoregressive transformer model undergoes a two-phase training process.

\textbf{Pre-training}: The AR model initially trains on large-scale audio-text paired datasets to learn fundamental audio representations.

\textbf{Fine-tuning}: Subsequently, the model fine-tunes on curated datasets with human-labeled text captions to enhance its musicality and adherence to prompts.

\textbf{Flow-matching Model Training}: The SRFM model trains using paired low- and high-resolution audio tokens to learn the upscaling transformation.

The multi-stage training regime ensures that each component of InspireMusic optimizes for its specific task while maintaining overall coherence in the final generated output.

In summary, InspireMusic represents a general and flexible framework for generating high-quality, long-form music. It combines the strengths of autoregressive transformers, audio tokenizers, and a super-resolution flow-matching model to produce $48kHz$ audio directly from text prompts. By leveraging multiple tokenizers and state-of-the-art LLMs like Qwen, the framework efficiently generates music with both high fidelity and structural coherence, pushing the boundaries of language model-based generative approaches.

\section{Dataset} \label{sec:data}
The pre-trained dataset includes weakly labeled audio and text. The audio part comprises about $29$ billion tokens of audio data (approximately $100,000$ hours). The audio data is resampled into $24kHz$ or $48kHz$ and segmented into $30$-second clips. Unless specified otherwise, the audio is downmixed to mono. The text part comprises around $740$ million tokens of text data obtained by a large language model.

The evaluation datasets utilized in this study include MusicCaps~(\cite{agostinelli2023musiclm}), Song Describer~(\cite{manco2023song}), and an in-house test dataset. In this version, the in-house test set comprises a collection of $30$ in-house music pieces with text and audio prompts. MusicCaps consists of $5521$ ten-second music samples, each accompanied by a free-text caption and a list of music aspects, such as genre, mood, tempo, and instrumentation, provided by expert musicians. Song Describer contains approximately $1100$ captions for $706$ permissively licensed music recordings. This dataset evaluates models on music and language tasks, including music captioning, text-to-music generation, and music-language retrieval.

\subsection{Audio Tokenization}
To enhance the efficiency of training the autoregressive model (AR), we convert the audio waveform into discrete audio tokens using WavTokenizer~(\cite{ji2024wavtokenizer}), a high-bitrate compression model that operates at a frame rate of $75Hz$. WavTokenizer employs a single quantizer layer to discretize audio into a sequence of one-dimensional tokens. Its encoder architecture mirrors that of EnCodec~(\cite{defossez2023encodec}), featuring a single quantizer layer, while its decoder draws inspiration from VOCOS~(\cite{siuzdak2024vocos}), incorporating an inverse Fourier transform (iFFT) structure to improve reconstruction quality. The model contains approximately $430M$ parameters. VOCOS functions as a neural vocoder that bridges the gap between time-domain and Fourier-based approaches. It generates spectral coefficients, facilitating rapid audio reconstruction through inverse Fourier transform.

We employ a non-causal HiFi-Codec model with $6$ layers for $48kHz$ monophonic audio, utilizing a stride of $320$ to achieve a frame rate of $150Hz$. The model comprises approximately $128M$ parameters. Embeddings are quantized using Residual Vector Quantization (RVQ) with four quantizers, each possessing a codebook size of $2048$. Following the methodology outlined by (\cite{yang2023hificodec}), we train the model on one-second audio segments randomly cropped from the audio waveforms.

\subsection{Text Captions}
\begin{figure}[ht]
    \centering
    \includegraphics[width=\linewidth]{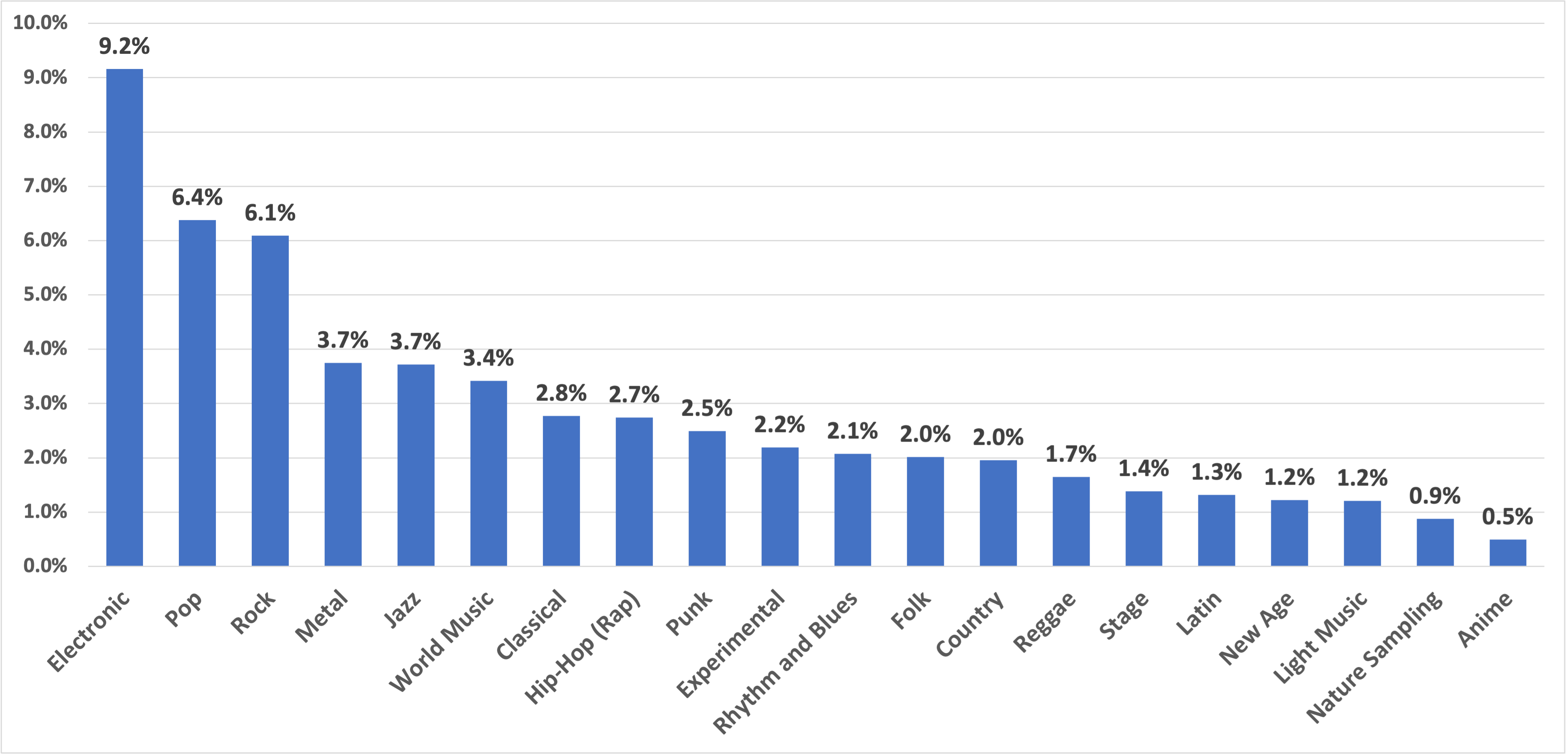}
    \caption{The statistical distribution of music genres in the pre-trained dataset.}
    \label{fig: stats}
\end{figure}

We utilize a robust large language model to generate text descriptions for music data based on tags, e.g., genres, instruments, styles, soundscapes, etc. The statistical distribution of the top $20$ music genres in the dataset appears in Figure~\ref{fig: stats}.

\vspace{-0.5cm}
\section{Experiments} \label{sec: experiment}
We conduct several subjective and objective experiments to evaluate InspireMusic against top-tier open-source models such as MusicGen~(\cite{musicgen}) and Stable Audio 2.0~(\cite{stableaudio2}). The evaluation of model performance in this section relies on an in-house test dataset unless otherwise specified.

\subsection{Experimental Setup}
Experiments take place on diverse datasets, including public benchmarks like MusicCaps~(\cite{agostinelli2023musiclm}), Song Describer~(\cite{manco2023song}), and proprietary in-house datasets covering various genres (e.g., electronic, classical, jazz, etc.). All models train on a cluster of H800 GPUs using the Adam optimizer with a base learning rate of $1\times10^{-4}$ and a scheduler that includes a warm-up learning step of $5000$. The GPU training hours of audio tokenizers, large language models, and flow-matching models are approximately $2352$, $4032$, $4704$, and $4032$, respectively. Parameter sizes of InspireMusic models appear in Table~\ref{tab: models}. The token size of InspireMusic-0.5B and InspireMusic-1.5B models is $156032$. The input dimension of WavTokenizer is $768$ and the output token size is $4096$. 

We train InspireMusic series models using a multi-stage training process. The training process includes pre-training, fine-tuning with $30$-second audio segmentations, and fine-tuning with full-length audio waveforms. In the first pre-training stage, the LLM model trains with audio tokens extracted via $75Hz$ WavTokenizer. In the second training stage, the LLM model trains with text captions alongside the corresponding audio tokens. In the third training stage, the LLM model trains with human-labeled text captions and audio tokens.

In the flow-matching training, $24kHz$ audio waveforms are extracted into discrete audio tokens as inputs, and the corresponding $48kHz$ audio waveforms are extracted into latent features via a $150Hz$ HiFi-Codec as the prediction outputs of the flow-matching model. WavTokenizer trains on a set of $24kHz$ music data, while HiFi-Codec trains on $48kHz$ music data.

\subsection{Evaluation}
In the experimental evaluation part, we evaluate InspireMusic models based on Qwen 2.5\footnote{https://github.com/QwenLM/Qwen2.5}, MusicGen-Small\footnote{https://huggingface.co/facebook/musicgen-small}, MusicGen-Medium\footnote{https://huggingface.co/facebook/musicgen-medium}, MusicGen-Large\footnote{https://huggingface.co/facebook/musicgen-large}, Stable Audio 2.0\footnote{https://www.stableaudio.com/} in in-house dataset as well as publicly available datasets including MusicCaps, Song Describer in terms of objective and subjective evaluation methods. 

\begin{table*}[t]
\centering
\caption{The objective evaluation of InspireMusic comparing with MusicGen-Small, MusicGen-Medium, MusicGen-Large and Stable Audio 2.0 on text-to-music task, in terms of KL$_{passt}$, FD$_{openl3}$, CLAP$_{score}$, respectively.} \label{tab:obj_t2m}
\footnotesize
\begin{tabular}{l|c|c|c}
\toprule
\multicolumn{1}{c|}{\multirow{2}{*}{Model}} & \multicolumn{3}{c}{Text to Music} 
     \\ 
                                         & \multicolumn{1}{c|}{KL$_{passt}$↓}      & \multicolumn{1}{c|}{FD$_{openl3}$↓}      & \multicolumn{1}{c}{CLAP$_{score}$↑} \\ \midrule
MusicGen-Small (300M)                  & \multicolumn{1}{c|}{0.822}              & \multicolumn{1}{c|}{201.253}             & 0.303                           \\ 
MusicGen-Medium (1.5B)                 & \multicolumn{1}{c|}{0.729}              & \multicolumn{1}{c|}{204.179}             & 0.312                           \\
MusicGen-Large (3.3B)                  & \multicolumn{1}{c|}{0.738}              & \multicolumn{1}{c|}{191.927}             & 0.304                           \\ 
StableAudio2.0 (1.1B)                  & \multicolumn{1}{c|}{0.576}              & \multicolumn{1}{c|}{76.737}              & \textbf{0.397}                  \\ \midrule
\begin{tabular}[c]{@{}l@{}}Cascade System\\ (InspireMusic-1.5B-Long w/o flow \\ + 48kHz super-resolution model)\end{tabular} & \multicolumn{1}{c|}{0.631}              & \multicolumn{1}{c|}{113.077}             & 0.246                           \\ \midrule
InspireMusic-1.5B-24kHz               & \multicolumn{1}{c|}{0.618}              & \multicolumn{1}{c|}{98.747}              & 0.258                           \\ 
InspireMusic-0.5B-24kHz               & \multicolumn{1}{c|}{0.790}              & \multicolumn{1}{c|}{94.023}              & 0.182                           \\ 
InspireMusic-0.5B                     & \multicolumn{1}{c|}{0.691}              & \multicolumn{1}{c|}{83.594}              & 0.246                           \\ 
InspireMusic-1.5B-Long w/o flow       & \multicolumn{1}{c|}{0.508}              & \multicolumn{1}{c|}{66.091}              & 0.264                           \\ 
InspireMusic-1.5B                     & \multicolumn{1}{c|}{0.782}              & \multicolumn{1}{c|}{65.162}              & 0.243                           \\ 
InspireMusic-1.5B-Long                & \multicolumn{1}{c|}{\textbf{0.378}}     & \multicolumn{1}{c|}{\textbf{63.429}}     & 0.324                           \\ \bottomrule
\end{tabular}
\end{table*}

\begin{table*}[t]
\centering
\caption{The objective evaluation of InspireMusic on music continuation task, in terms of KL$_{passt}$, FD$_{openl3}$, CLAP$_{score}$, respectively.} \label{tab:obj_con}
\begin{tabular}{l|ccc}
\toprule
\multicolumn{1}{c|}{\multirow{2}{*}{Model}} & \multicolumn{3}{c}{Music Continuation}                                                              \\
\multicolumn{1}{c|}{}                       & \multicolumn{1}{l|}{KL$_{passt}$↓}      & \multicolumn{1}{l|}{FD$_{openl3}$↓}      & \multicolumn{1}{c}{CLAP$_{score}$↑} \\ \midrule
InspireMusic-0.5B                            & \multicolumn{1}{c|}{0.515}          & 67.949                           & \textbf{0.273}                      \\ 
InspireMusic-0.5B w/o flow                 & \multicolumn{1}{c|}{0.348}          & \textbf{60.966}                  & 0.264                           \\ 
InspireMusic-1.5B                            & \multicolumn{1}{c|}{0.346}          & 321.833                          & 0.246                           \\ 
InspireMusic-1.5B-Long w/o flow            & \multicolumn{1}{c|}{\textbf{0.228}} & 86.519                           & 0.264                           \\
InspireMusic-1.5B-Long                       & \multicolumn{1}{c|}{0.411}          & 64.206                           & 0.257                           \\ \bottomrule
\end{tabular}
\end{table*}

\begin{table*}[ht]
\centering
\caption{The objective evaluation of InspireMusic comparing with MusicGen-Small, MusicGen-Medium, MusicGen-Large, on \textbf{MusicCaps dataset} for the text-to-music task, in terms of KL$_{passt}$, FD$_{openl3}$, CLAP$_{score}$, respectively.} \label{tab:obj_musiccaps_t2m}
\begin{tabular}{l|ccc}
\toprule
\multirow{2}{*}{Model} & \multicolumn{3}{c}{MusicCaps}                                                                               \\  
                       & \multicolumn{1}{c|}{KL$_{passt}$↓}      & \multicolumn{1}{c|}{FD$_{openl3}$ ↓}     & \multicolumn{1}{c}{CLAP$_{score}$↑} \\ \midrule
MusicGen-Small (300M)         & \multicolumn{1}{c|}{1.061}          & \multicolumn{1}{c|}{325.156}              & 0.275                           \\ 
MusicGen-Medium (1.5B)       & \multicolumn{1}{c|}{0.999}          & \multicolumn{1}{c|}{316.787}              & \textbf{0.284}                   \\ 
MusicGen-Large (3.3B)        & \multicolumn{1}{c|}{0.975}          & \multicolumn{1}{c|}{317.254}              & 0.283                           \\ \midrule
InspireMusic-0.5B      & \multicolumn{1}{c|}{1.511}          & \multicolumn{1}{c|}{107.942}              & 0.145                           \\ 
InspireMusic-1.5B      & \multicolumn{1}{c|}{\textbf{0.610}} & \multicolumn{1}{c|}{\textbf{56.208}}     & 0.215                           \\ 
InspireMusic-1.5B-Long & \multicolumn{1}{c|}{0.708}          & \multicolumn{1}{c|}{101.020}              & 0.173                           \\ \bottomrule
\end{tabular}
\end{table*}

\begin{table*}[ht]
\centering
\caption{The objective evaluation of InspireMusic comparing with MusicGen-Small, MusicGen-Medium, MusicGen-Large, on \textbf{Song Describer dataset} for the text-to-music task, in terms of KL$_{passt}$, FD$_{openl3}$, CLAP$_{score}$, respectively.} \label{tab:obj_song_describer_t2m}
\begin{tabular}{l|ccc}
\toprule
\multicolumn{1}{c|}{\multirow{2}{*}{Model}} & \multicolumn{3}{c}{Song Describer}                                                                   \\ 
                       & \multicolumn{1}{c|}{KL$_{passt}$↓}      & \multicolumn{1}{c|}{FD$_{openl3}$↓}     & \multicolumn{1}{c}{CLAP$_{score}$↑} \\ \midrule
MusicGen-Small (300M)        & \multicolumn{1}{c|}{0.48}          & \multicolumn{1}{c|}{389.74}              & 0.24                           \\ 
MusicGen-Medium (1.5B)       & \multicolumn{1}{c|}{0.44}          & \multicolumn{1}{c|}{392.93}              & 0.24                           \\ 
MusicGen-Large  (3.3B)       & \multicolumn{1}{c|}{0.40}          & \multicolumn{1}{c|}{373.33}              & 0.26                           \\ \midrule
MusicGen-Large-Stereo (3.3B) & \multicolumn{1}{c|}{0.50}          & \multicolumn{1}{c|}{213.76}              & 0.28                           \\ 
StableAudio2.0  (1.1B)       & \multicolumn{1}{c|}{0.34}          & \multicolumn{1}{c|}{89.33}               & \textbf{0.39}                  \\ \midrule
InspireMusic-1.5B-Long       & \multicolumn{1}{c|}{\textbf{0.23}} & \multicolumn{1}{c|}{\textbf{86.55}}     & 0.29                           \\ \bottomrule
\end{tabular}
\end{table*}

\begin{table*}[t]
\centering
\caption{Results of subjective listening test for the \textbf{Text-to-Music} task, in terms of CMOS score with the range of $1.0\,\sim\,5.0$, reporting with both mean and CI95 scores.} \label{tab:subj_cmos_t2m}
\footnotesize
\begin{tabular}{l|llll}
\toprule
Model                                    & \multicolumn{1}{c|}{Alignment↑}          & \multicolumn{1}{c|}{Audio Quality↑}      & \multicolumn{1}{c|}{Musicality↑}         & \multicolumn{1}{c}{Overall↑} \\ \midrule
InspireMusic-0.5B                        & \multicolumn{1}{l|}{2.88 ± 0.60}           & \multicolumn{1}{l|}{2.86 ± 0.63}           & \multicolumn{1}{l|}{2.85 ± 0.68}           & 2.87 ± 0.64                     \\ 
InspireMusic-0.5B-24kHz                  & \multicolumn{1}{l|}{3.06 ± 0.64}           & \multicolumn{1}{l|}{3.03 ± 0.51}          & \multicolumn{1}{l|}{3.01 ± 0.63}           & 3.08 ± 0.56                     \\ 
InspireMusic-0.5B   (w/o flow)           & \multicolumn{1}{l|}{3.08 ± 0.48}           & \multicolumn{1}{l|}{3.08 ± 0.54}           & \multicolumn{1}{l|}{3.09 ± 0.57}           & 3.09 ± 0.51                     \\ 
InspireMusic-1.5B-24kHz                  & \multicolumn{1}{l|}{2.93 ± 0.56}           & \multicolumn{1}{l|}{2.88 ± 0.51}          & \multicolumn{1}{l|}{2.98 ± 0.61}           & 2.88 ± 0.58                    \\ 
InspireMusic-1.5B                        & \multicolumn{1}{l|}{2.90 ± 0.64}          & \multicolumn{1}{l|}{2.81 ± 0.75}           & \multicolumn{1}{l|}{2.81 ± 0.74}           & 2.89 ± 0.65                    \\ 
InspireMusic-1.5B   (w/o flow)           & \multicolumn{1}{l|}{3.01 ± 0.53}          & \multicolumn{1}{l|}{2.96 ± 0.60}          & \multicolumn{1}{l|}{3.00 ± 0.67}          & 3.02 ± 0.58                     \\ 
InspireMusic-1.5B-Long   (w/o flow)      & \multicolumn{1}{l|}{3.26 ± 0.53}           & \multicolumn{1}{l|}{3.17 ± 0.60}           & \multicolumn{1}{l|}{3.15 ± 0.63}           & 3.28 ± 0.56                    \\
\textbf{InspireMusic-1.5B-Long}          & \multicolumn{1}{l|}{\textbf{3.27 ± 0.69}} & \multicolumn{1}{l|}{\textbf{3.24 ± 0.65}} & \multicolumn{1}{l|}{\textbf{3.29 ± 0.63}} & \textbf{3.34 ± 0.60}            \\ \midrule
MusicGen-Small                           & \multicolumn{1}{l|}{3.00 ± 0.54}          & \multicolumn{1}{l|}{2.95 ± 0.58}          & \multicolumn{1}{l|}{2.97 ± 0.61}           & 3.02 ± 0.55                    \\ 
MusicGen-Medium                          & \multicolumn{1}{l|}{2.90 ± 0.47}          & \multicolumn{1}{l|}{2.89 ± 0.48}          & \multicolumn{1}{l|}{2.81 ± 0.64}          & 2.86 ± 0.51                    \\ 
MusicGen-Large                           & \multicolumn{1}{l|}{2.94 ± 0.49}           & \multicolumn{1}{l|}{2.92 ± 0.53}           & \multicolumn{1}{l|}{2.95 ± 0.55}          & 2.99 ± 0.49                     \\ 
StableAudio2.0                           & \multicolumn{1}{l|}{3.10 ± 0.67}          & \multicolumn{1}{l|}{3.03 ± 0.62}          & \multicolumn{1}{l|}{3.10 ± 0.67}          & 3.11 ± 0.68                    \\ \bottomrule
\end{tabular}
\end{table*}

\begin{table*}[ht]
\centering
\caption{Results of the subjective listening test for \textbf{Music Continuation} task, in terms of CMOS score with the range of $1.0\,\sim\,5.0$, reporting with both mean and CI95 scores.} \label{tab:subj_cmos_con}
\footnotesize
\begin{tabular}{l|llll}
\toprule                                                                                                                  
Models                              & \multicolumn{1}{c|}{Alignment↑}        & \multicolumn{1}{c|}{Audio Quality↑}      & \multicolumn{1}{c|}{Musicality↑}        & \multicolumn{1}{c}{Overall↑} \\ \midrule
InspireMusic-0.5B                   & \multicolumn{1}{l|}{2.98 ± 0.71}        & \multicolumn{1}{l|}{2.98 ± 0.78}          & \multicolumn{1}{l|}{3.00 ± 0.80}         & 3.01 ± 0.73                    \\ 
InspireMusic-0.5B-24kHz             & \multicolumn{1}{l|}{3.07 ± 0.70}        & \multicolumn{1}{l|}{2.94 ± 0.79}           & \multicolumn{1}{l|}{2.98 ± 0.86}          & 3.04 ± 0.79                     \\ 
InspireMusic-0.5B (w/o flow)      & \multicolumn{1}{l|}{3.20 ± 0.64}          & \multicolumn{1}{l|}{3.18 ± 0.64}           & \multicolumn{1}{l|}{3.12 ± 0.66}         & 3.20 ± 0.61                    \\ 
InspireMusic-1.5B-24kHz             & \multicolumn{1}{l|}{3.02 ± 0.72}         & \multicolumn{1}{l|}{2.89 ± 0.73}          & \multicolumn{1}{l|}{2.97 ± 0.81}          & 2.98 ± 0.77                    \\ 
InspireMusic-1.5B                   & \multicolumn{1}{l|}{3.01 ± 0.68}        & \multicolumn{1}{l|}{2.87 ± 0.83}           & \multicolumn{1}{l|}{2.88 ± 0.88}         & 3.01 ± 0.80                    \\ 
InspireMusic-1.5B (w/o flow)      & \multicolumn{1}{l|}{2.96 ± 0.61}         & \multicolumn{1}{l|}{2.94 ± 0.71}           & \multicolumn{1}{l|}{2.90 ± 0.78}         & 2.96 ± 0.67                    \\ 
InspireMusic-1.5B-Long (w/o flow) & \multicolumn{1}{l|}{3.11 ± 0.63}        & \multicolumn{1}{l|}{3.07 ± 0.61}          & \multicolumn{1}{l|}{3.03 ± 0.67}          & 3.16 ± 0.64                     \\ 
\textbf{InspireMusic-1.5B-Long}              & \multicolumn{1}{l|}{\textbf{3.30 ± 0.61}} & \multicolumn{1}{l|}{\textbf{3.22 ± 0.64}} & \multicolumn{1}{l|}{\textbf{3.17 ± 0.66}} & \textbf{3.21 ± 0.58}            \\ \bottomrule
\end{tabular}
\end{table*}

\begin{table*}[ht]
\centering
\caption{The ablation study of the proposed framework, take InspireMusic-1.5B-Long as an example, without flow-matching or super-resolution on text-to-music task, in the aspect of subjective evaluation, in terms of alignment, audio quality, musicality, and overall performance, respectively.} \label{tab:subjective_fm_t2m}
\begin{tabular}{l|cccc}
\toprule                                                                                             
Model                  & \multicolumn{1}{c|}{Alignment↑} & \multicolumn{1}{c|}{Audio Quality↑} & \multicolumn{1}{c|}{Musicality↑} & \multicolumn{1}{c}{Overall↑} \\ \midrule
InspireMusic-1.5B-Long & \multicolumn{1}{c|}{\textbf{3.27 ± 0.69}}  & \multicolumn{1}{c|}{\textbf{3.24 ± 0.65}}      & \multicolumn{1}{c|}{\textbf{3.29 ± 0.63}}   & \textbf{3.34 ± 0.60}                     \\ 
~~~~w/o flow-matching                  & \multicolumn{1}{c|}{3.26 ± 0.53}  & \multicolumn{1}{c|}{3.17 ± 0.60}      & \multicolumn{1}{c|}{3.15 ± 0.63}   & 3.28 ± 0.56                     \\ 
~~~~w/o Super-Resolution                & \multicolumn{1}{c|}{3.09 ± 0.62}  & \multicolumn{1}{c|}{3.02 ± 0.61}      & \multicolumn{1}{c|}{3.09 ± 0.64}   & 3.11 ± 0.57                     \\ \bottomrule
\end{tabular}
\end{table*}

\begin{table*}[ht]
\centering
\caption{The ablation study of the proposed framework, e.g., InspireMusic-1.5B-Long, without flow-matching or super-resolution on music continuation task, in the aspect of subjective evaluation, in terms of alignment, audio quality, musicality, and overall performance, respectively.} \label{tab:subjective_fm_con}
\begin{tabular}{l|c|c|c|c}
\toprule
Model                  & \multicolumn{1}{c|}{Alignment↑} & \multicolumn{1}{c|}{Audio Quality↑} & \multicolumn{1}{c|}{Musicality↑} & \multicolumn{1}{c}{Overall↑} \\ \midrule
InspireMusic-1.5B-Long & \textbf{3.30 ± 0.61}               & \textbf{3.22 ± 0.64}                   & \textbf{3.17 ± 0.66}               & \textbf{3.21 ± 0.58}            \\ 
~~~~- w/o flow-matching      & 3.11 ± 0.63                      & 3.07 ± 0.61                            & 3.03 ± 0.67                        & 3.16 ± 0.64                     \\ 
~~~~- w/o Super-Resolution   & 3.23 ± 0.57                       & 3.16 ± 0.67                             & 3.17 ± 0.72                        & 3.24 ± 0.56                     \\ \bottomrule
\end{tabular}
\end{table*}

\begin{table*}[ht]
\centering
\caption{The objective evaluation of InspireMusic-1.5B-Long with and without SRFM under different CFG values on text-to-music and music continuation tasks in terms of KL$_{passt}$ and FD$_{openl3}$, CLAP$_{score}$, respectively.} \label{tab:ablation_cfg}
\scriptsize
\begin{tabular}{l|c|ccc|ccc}
\toprule
\multirow{2}{*}{Model}                           & \multicolumn{1}{c|}{\multirow{2}{*}{CFG}} & \multicolumn{3}{c|}{Text-to-Music}                                                                           & \multicolumn{3}{c}{Music Continuation}                                                                      \\ 
                                                 & \multicolumn{1}{c|}{}                           & \multicolumn{1}{l|}{KL$_{passt}$↓}      & \multicolumn{1}{l|}{FD$_{openl3}$↓}   & \multicolumn{1}{l|}{CLAP$_{score}$↑}     & \multicolumn{1}{l|}{KL$_{passt}$↓}      & \multicolumn{1}{l|}{FD$_{openl3}$↓} & \multicolumn{1}{l}{CLAP$_{score}$↑} \\ \midrule
  \multirow{5}{*}{w/ SRFM}                                                         & 3.0                                      & \multicolumn{1}{c|}{0.378}          & \multicolumn{1}{c|}{\textbf{63.429}}         & \multicolumn{1}{c|}{\textbf{0.324}}                   & \multicolumn{1}{c|}{\textbf{0.401}} & \multicolumn{1}{c|}{77.086}        & \multicolumn{1}{c}{0.286}                    \\  
                                                 & 5.0                                                 & \multicolumn{1}{c|}{\textbf{0.358}} &  \multicolumn{1}{c|}{101.281}   & \multicolumn{1}{c|}{0.238}                                & \multicolumn{1}{c|}{0.546}          & \multicolumn{1}{c|}{83.717}        & \multicolumn{1}{c}{0.219}                    \\  
                                                 & 7.0                                           & \multicolumn{1}{c|}{0.590}          & \multicolumn{1}{c|}{75.063}           & \multicolumn{1}{c}{0.272}                               & \multicolumn{1}{c|}{0.597}          & \multicolumn{1}{c|}{77.115}        & \multicolumn{1}{c}{0.217}                    \\ 
                                                 & 10.0                                          & \multicolumn{1}{c|}{0.599}          & 71.384            & \multicolumn{1}{c|}{0.255}                               & \multicolumn{1}{c|}{0.462}          & \textbf{75.698}     & \multicolumn{1}{c}{0.262}             \\ \midrule
\multirow{5}{*}{w/o SRFM}                                               & 3.0                                               & \multicolumn{1}{c|}{0.528}          & \multicolumn{1}{c|}{90.130}                 & \multicolumn{1}{c|}{0.270}          & \multicolumn{1}{c|}{\textbf{0.228}} & \multicolumn{1}{c|}{86.532}            & \multicolumn{1}{c}{0.287}                           \\ 
                                                 & 5.0                                                 & \multicolumn{1}{c|}{0.872}          & \multicolumn{1}{c|}{76.700}           & \multicolumn{1}{c|}{\textbf{0.288}}                & \multicolumn{1}{c|}{0.359}          & \multicolumn{1}{c|}{\textbf{57.565}}           & \multicolumn{1}{c}{\textbf{0.289}}        \\ 
                                                 & 7.0                                              & \multicolumn{1}{c|}{0.508}          & \textbf{66.091}          & \multicolumn{1}{c|}{0.264}                    & \multicolumn{1}{c|}{0.554}          & \multicolumn{1}{c|}{66.145}             & \multicolumn{1}{c}{0.225}               \\ 
                                                 & 10.0                                              & \multicolumn{1}{c|}{0.551}          & \multicolumn{1}{c|}{66.592}          & \multicolumn{1}{c|}{0.279}                             & \multicolumn{1}{c|}{0.428}          & \multicolumn{1}{c|}{59.373}           & \multicolumn{1}{c}{0.250}                \\ \bottomrule
\end{tabular}
\end{table*}

\begin{table*}[t]
\centering
\caption{The performance comparison of InspireMusic with different generation lengths (i.e., $30s$, $1min$, $5min$) on text-to-music task, comparing with Stable Audio 2.0.} \label{tab:ablation_length_con}
\footnotesize
\begin{tabular}{l|c|ccc}
\toprule
\multirow{2}{*}{Model}                  & \multirow{2}{*}{\begin{tabular}[c]{@{}c@{}}Generation \\ Audio Length\end{tabular}} & \multicolumn{3}{c}{Text-to-Music Task}                                                     \\ 
                                        &                                                                                           & \multicolumn{1}{c|}{KL$_{passt}$↓}      &\multicolumn{1}{c}{FD$_{openl3}$} ↓    & \multicolumn{1}{c}{CLAP$_{score}$↑}  \\ \midrule
\multirow{3}{*}{InspireMusic-1.5B-Long} & $30s$                                                                                           & \multicolumn{1}{c|}{0.378}          & \multicolumn{1}{c|}{63.429}     & \multicolumn{1}{c}{0.324}     \\ 
                                        & $1min$                                                                                         & \multicolumn{1}{c|}{0.378}            & \multicolumn{1}{c|}{69.800}     & \multicolumn{1}{c}{0.266}      \\  
                                        & $5min$                                                                                         & \multicolumn{1}{c|}{0.391} &  \multicolumn{1}{c|}{66.309} & \multicolumn{1}{c}{0.272} \\ \midrule
Stable Audio 2.0                      & $3min$                                                                                 & \multicolumn{1}{c|}{0.576}          & \multicolumn{1}{c|}{76.737}   & \multicolumn{1}{c}{0.397}       \\ \bottomrule
\end{tabular}
\end{table*}
\subsection{Evaluation Metrics}
We evaluate the proposed approach using objective and subjective metrics. For objective evaluation, we use three metrics: the Fréchet Distance (FD), the Kullback-Leibler Divergence (KL), and the CLAP score. For subjective evaluation, we use four metrics: audio-text alignment, audio quality, musicality, and overall performance. The detailed evaluation metrics used to assess performance are listed below.
\vspace{-0.5cm}
\subsubsection{Objective Evaluations}
\textbf{FD${openl3}$}: The Fr'{e}chet Distance (FD) evaluates the similarity between the statistics of a generated audio set and a reference audio set in a feature space. A low FD${openl3}$ score indicates that the generated audio is plausible and closely matches the reference.

\textbf{KL$_{passt}$}: We use PaSST~(\cite{koutini2021efficient}), a state-of-the-art audio tagger trained on AudioSet~(\cite{gemmeke2017audioset}), to compute the Kullback–Leibler (KL) divergence~(\cite{kullback1951information}) over the probabilities of the labels between the generated and the reference audio. It computes the KL divergence over the probabilities of the labels between the original and the generated music. The generated music is expected to share concepts similar to the reference music when the KL is low.

\textbf{CLAP$_{score}$}: The CLAP score~\cite{elizalde2022claplearningaudioconcepts} is computed between the music description and the generated audio to quantify audio-text alignment, using the official pre-trained CLAP model.
\vspace{-0.5cm}
\subsubsection{Subjective Evaluations}
For subjective evaluation, we use the Comparative Mean Opinion Score (CMOS), based on feedback from 10 raters with professional backgrounds in music. We define the following four dimensions for assessment.

\textbf{Audio-Text Alignment}: Performance is assessed by measuring the alignment between textual prompts and the generated musical output using cross-modal retrieval metrics. Text alignment evaluates how closely the generated audio corresponds to the input text or audio prompts, ensuring that the produced music aligns with the provided text prompts. Raters assign CMOS scores based on this alignment. In the audio-text alignment test, raters evaluate the correspondence between audio and text on a scale of 1.0 to 5.0.

\textbf{Audio Quality}: We measure audio quality using Fr'{e}chet Audio Distance (FD), which quantifies the perceptual similarity between generated audio and real music. Audio quality assesses whether the generated audio is of low fidelity (i.e., with artifacts, distortion, and noise) or high fidelity. Evaluators assign a CMOS based on musical coherence. In text-to-music tasks, this involves determining whether the entire generated piece maintains consistency and coherence in style, genre, and structure. For music continuation tasks, it assesses whether the generated music follows the melody, rhythm, and type of the music prompt, ensuring coherence.

\textbf{Musicality}: Musicality evaluates musical attributes, including the originality of melodies and rhythms, effective use of harmony, adherence to idiomatic musical forms (such as genre and style), structural coherence, appropriate chord progressions, distinctive rhythmic patterns, well-balanced instrumentation, and the creation of an appropriate ambiance and atmosphere. We evaluate musicality using Comparative Mean Opinion Score (CMOS) tests, where musically trained raters assess the harmonic, rhythmic, and structural quality of the generated outputs. 

\textbf{Overall Performance}: We evaluate overall performance through both objective analysis and automatic metrics that capture long-term dependencies and musical form. For the overall quality test, raters rate the perceptual quality of the provided samples on a scale from 1.0 to 5.0.
\subsection{Tasks}
InspireMusic models are evaluated on text-to-music and music continuation tasks, respectively.

\textbf{Text-to-Music}: Generate high-fidelity music with long-form coherence from the input text prompts.

\textbf{Music Continuation}: Continue the music composition based on the input audio prompts.
\vspace{-0.5cm}
\subsection{Results} \label{sec:results}
In this section, the default setting of InspireMusic is an autoregressive transformer with a super-resolution flow-matching model. The InspireMusic series model without flow matching denotes we use the WavTokenizer decoder to transform generated tokens into waveform directly. A model without super-resolution means the model uses a $24kHz$ flow matching model to generate $24kHz$ audio waveform instead of using SRFM to generate $48kHz$ audio.

The objective evaluation of InspireMusic compared with MusicGen and Stable Audio 2.0 concerning the text-to-music and music continuation tasks appears in Table~\ref{tab:obj_t2m} and Table~\ref{tab:obj_con}, respectively. Our experiments demonstrate that the InspireMusic-1.5B-Long model outperforms MusicGen-Small, MusicGen-Medium, MusicGen-Large, and Stable Audio 2.0 across all evaluation dimensions. In subjective evaluations for the text-to-music task, InspireMusic-1.5B-Long achieves a CMOS score that is 7\% higher relative to Stable Audio 2.0 and shows a 14\% improvement over InspireMusic-0.5B. Additionally, InspireMusic-1.5B-Long surpasses InspireMusic-0.5B by 6.5\% in CMOS score for the same task. Objective metrics, including lower FD and improved KL divergence scores, further confirm the superior audio quality and structural coherence of InspireMusic.

Table~\ref{tab:obj_musiccaps_t2m} presents the comparison of InspireMusic with MusicGen on the MusicCaps test set for the text-to-music task. The InspireMusic-1.5B model performs better than the MusicGen models on KL and FD scores. When considering the CLAP score, it is important to note that this version of InspireMusic models does not train with AudioCaps or AudioSet data. In contrast, both the CLAP model and the models used by Stable Audio 2.0 and MusicGen utilize text and audio alignment or embeddings derived from these datasets, which raises potential concerns regarding information leakage. Furthermore, the text captions for InspireMusic are generated using a large language model, leading to a stylistic approach that differs from the captions produced by MusicCaps or Song Describer. Consequently, the CLAP score for InspireMusic remains lower than that of the MusicGen models and Stable Audio 2.0.

Table~\ref{tab:obj_song_describer_t2m} displays the objective evaluation results of InspireMusic-1.5B-Long without the flow-matching model compared with MusicGen and Stable Audio 2.0 on the Song Describer dataset. The results show that InspireMusic outperforms both models in terms of KL and FD scores. In terms of CLAP score, InspireMusic has better performance than MusicGen models, but lower than Stable Audio 2.0 due to the same reason above.

Table~\ref{tab:subj_cmos_t2m} and Table~\ref{tab:subj_cmos_con} demonstrate the subjective listening test results of InspireMusic comparing with MusicGen and Stable Audio 2.0 models in terms of CMOS score with the range of 1.0 and 5.0. The results in this table are reported with both mean with confidence intervals of 95\% (CI95). The InspireMusic-1.5B-Long could generate longer music than InspireMusic-1.5B, where the generated tokens may include more errors, with the super-resolution flow-matching model. The flow-matching model could help to correct some artifacts within the generated audio tokens. We have done some tests that show the output waveform of the super-resolution flow-matching model has better performance than using the WavTokenizer decoder to transform generated tokens to audio waveforms directly (i.e., w/o FM). Thus, InspireMusic-1.5B-Long with flow-matching performs better than InspireMusic-1.5B with or without flow-matching.

For example, in tests involving extended compositions, InspireMusic successfully generates a 5-minute piece with clearly defined musical structures (i.e., intro, verse, chorus, outro) and minimal artifacts. In contrast, MusicGen often exhibits repetition and loss of structure after approximately 30 seconds. Furthermore, the high-resolution outputs produced by the SRFM module result in significantly clearer audio compared to Stable Audio 2.0, as validated by both KL, FD scores, and listening tests.

\subsection{Ablation Studies}
To assess the contributions of each component, we conduct ablation studies, including the evaluation of the models with or without the flow-matching model, with or without super-resolution.

Table~\ref{tab:subjective_fm_t2m} and Table~\ref{tab:subjective_fm_con} present the subjective performance comparison of the InspireMusic-1.5B-Long with or without flow-matching, without super-resolution on text-to-music and music continuation tasks, respectively, in terms of CMOS score in the aspects of alignment, audio quality, musicality, and overall performance. The default setting for InspireMusic-1.5B-Long is with the SRFM model. The evaluation results show that removing the SRFM model results in a notable drop in audio fidelity, highlighting its importance in achieving $48kHz$ quality. Using only the WavTokenizer decoder (without the Hifi-Codec vocoder) leads to reduced musical detail and a loss in dynamic range. The autoregressive language model without SRFM reduces long-term coherence, particularly in extended compositions.

Table~\ref{tab:ablation_cfg} presents the objective evaluation of the InspireMusic-1.5B-Long model with and without SRFM under different CFG values on text-to-music as well as music continuation tasks, in terms of CLAP${score}$, KL${passt}$, and FD$_{openl3}$.
Table~\ref{tab:ablation_length_con} presents the objective evaluation results of the InspireMusic-1.5B-Long model under different audio generation lengths on both text-to-music and music continuation tasks. InspireMusic-1.5B-Long, which generates an audio length of 5 minutes, outperforms Stable Audio 2.0, which generates audio for 3 minutes, in terms of KL and FD scores.

\vspace{-0.5cm}
\section{Conclusion} \label{sec:conclusion} 
In this paper, we introduce InspireMusic, a unified framework that integrates an autoregressive transformer model with super-resolution flow-matching to generate long-form and high-fidelity music. Our approach leverages advanced audio tokenizers, such as WavTokenizer and Hifi-Codec, to extract audio representations at various sampling rates. Utilizing Qwen 2.5 as the backbone large language model, we train an autoregressive transformer to generate audio tokens. The super-resolution flow-matching component then maps audio tokens at a lower token rate, which contain both semantic and acoustic information, into fine-grained latent acoustic features derived from an acoustic codec at a higher token rate, and then decodes these features into high-fidelity audio with a higher sampling rate. We demonstrate that InspireMusic produces good quality long-form music for both text-to-music and music continuation tasks, controlled by diverse text and audio prompts. Evaluation results show that InspireMusic has comparable performance with the current top-tier open-source models such as the MusicGen series and Stable Audio 2.0 on both objective and subjective metrics. This work provides an efficient and scalable pathway for generating diverse, long-form, coherent, and high-quality audio compositions.

\section{Authors}

\begin{multicols}{3} 
Chong Zhang$^{\star}$ \\ Yukun Ma$^{\star}$ \\ Qian Chen \\ Wen Wang \\ Shengkui Zhao \\Zexu Pan \\ Hao Wang \\ Chongjia Ni \\ Trung Hieu Nguyen \\ Kun Zhou \\ Yidi Jiang \\Chaohong Tan \\ Zhifu Gao \\ Zhihao Du \\ Bin Ma \\ 
\end{multicols} 
\textsuperscript{$\star$} Equal contribution.

\section{Acknowledgments}
We express our gratitude to Dianwen Ng, Jiaqi Yip, Lingyun Zuo, Changfeng Gao, Nan Zhao, Shiliang Zhang, and Zhijie Yan for their help.
\bibliography{InspireMusic}
\bibliographystyle{InspireMusic}
\appendix
\end{document}